\documentclass[12pt]{iopart}

\usepackage{iopams}
\usepackage{graphicx}

\usepackage{epstopdf}
\newcommand{\ZN}{\mathbb{Z}_N}
\newcommand{\wrap}[1]{\left({#1}\right)}
\newcommand{\ii}{\rmi} 
\newcommand{\ee}{\rme}
\newcommand{\graphical}[2][]
{\vcenter{\hbox{\includegraphics[{#1}]{#2}}}}

\begin{document}

\title{Integrability as a consequence of discrete holomorphicity: 
loop models}

\author{I T Alam$^1$ and M T Batchelor$^{2,1,3}$}

\address{$^1$ Department of Theoretical Physics, 
Research School of Physics and Engineering, 
The Australian National University, Canberra ACT 0200, Australia}
\address{$^2$ Centre for Modern Physics, Chongqing University,
Chongqing 400044, China}
\address{$^3$ Mathematical Sciences Institute, 
The Australian National University, Canberra ACT 0200, Australia}

\ead{Imam.Alam@anu.edu.au}
\ead{Murray.Batchelor@anu.edu.au}

%\maketitle

\begin{abstract}
In this paper, we explore the relationship between integrability and
the discrete holomorphicity of a class of complex lattice observables 
in the context of the Potts dense loop model and the $O(n)$ dilute 
loop model. It is shown that the conditions for integrability, namely,
the inversion and Yang-Baxter relations, can be derived from the
condition of holomorphicity of the observables. Furthermore, the 
$Z$-invariance of the models is shown to result in the invariance of
the observables on the boundary of a sublattice under reshuffling of
the rhombuses of its planar rhombic embedding.
\end{abstract}

\section{Introduction}
The celebrated rigorous proof of conformal invariance
of the Ising model \cite{SS-CIRCM, SS-DCP} spurred renewed interest 
in the application of discrete complex analysis to two-dimensional 
critical lattice models. The approach has inspired mathematical 
proofs for a number of notable results in statistical physics.
For the Ising model, these include 
the conformal invariance of spin correlations \cite{CHI-CISC} and
the energy density \cite{HS-EDPIM} and the existence of the
scaling limit of domain walls \cite{CS-UIM}. For self-avoiding 
walks on the hexagonal lattice, proofs have been given for the 
connective constant \cite{DCS-CCHL} and the critical fugacity of surface 
adsorption \cite{BDG-CFSA,Beaton}.

Of course, the solvability of these problems ought, ultimately, to be
ascribed to the integrability \cite{B-ESMSM} of the models. In a
parallel development, it was discovered that a weakening of the
condition of discrete holomorphicity, namely, imposing only the
condition of vanishing discrete contour sum, is obeyed by a class of
parafermionic observables on the lattice for a wide variety of models
such as the Potts \cite{RC-HPPM}, the $\ZN$ \cite{RC-DHPL}
and the $O(n)$ loop \cite{IC-DHPILM} models, both in the 
bulk \cite{RC-HPPM,RC-DHPL,IC-DHPILM} and with a 
boundary \cite{I,Lee}.
Intriguingly, for these observables, the condition holds only on 
the integrable critical manifold of the Boltzmann weights,
warranting a deeper understanding of the connection between
integrability and holomorphicity. In this paper, we refer to this
weak form of holomorphicity only as discrete holomorphicity for
 simplicity.

Subsequently, we showed \cite{AB-ICDH} that for 
Fateev-Zamolodchikov self-dual $\ZN$ models \cite{FZ-SDSSTR},
discrete holomorphicity necessarily implies Yang-Baxter integrability
conditions. In this paper, we demonstrate the generality of the 
connection by showing that analogous arguments hold for the loop
models. The integrable structure in the cases considered here 
can alternatively be understood as arising from the quantum group
\cite{D-QG} structure.
In \cite{IWWZ-DHQAA}, the relevant parafermions
were identified as non-local conserved currents \cite{BF-QGSQFT}
corresponding to this symmetry with their conservation law given
by the condition of discrete holomorphicity. 
This approach leads directly to the Yang-Baxter equation by 
exploiting the associativity of the algebra. 
Our approach here,
however, is focused on the connection with the $Z$-invariance
\cite{B-S8V} of these models.
The holomorphic
observable, remarkably, also has close connections \cite{L-FOIM} 
to the combinatorial approach \cite{KW-CSIM}  for the special case of 
the Ising model. 

In \sref{sec:dense} we present our results for the critical Potts
model in its dense loop formulation, that is, we derive the inversion
relation and the Yang-Baxter equation from the condition of
holomorphicity, and in \sref{sec:dilute} we extend
the results to the $O(n)$ dilute loop model. In \sref{sec:invariance} we
present an alternative characterization of the observables which
uses the $Z$-invariance of the model to define the observables on the
boundary for the whole equivalence class of Baxter lattices sharing
that boundary. Some concluding remarks are given in section 5.

\section{The dense loop model}\label{sec:dense}

\subsection{Yang-Baxter integrability}
It is well-known \cite{FK-RCM, BKW-EPM} that the $N$-state critical
Potts model on a planar graph can be mapped to a dense loop model
defined on its medial graph. We call a face of the covering graph 
(the planar dual of the medial graph) a plaquette and the edge of the 
original graph on which the plaquette is placed its axis. Each closed 
loop gets a statistical weight
\[
n = \sqrt{N} = 2\,\cos\lambda \quad\quad 
(0\le N\le 4, 
\quad 0 \le \lambda \le \frac{\pi}{2})
\]
and each plaquette can have on it one of two configurations with 
weights $a$ and $b$ that depend on the interaction energy of its axis.
We represent the $R$-matrix graphically as a formal linear 
combination of the configurations, or the connectivities,
on the plaquette,
which are essentially the 2 possible pairings between the 4 midpoints
of the sides, 
\begin{equation*}
\graphical{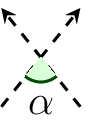}
\;=\; a_\alpha\;\;
\graphical[scale=1.25]{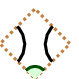}\;+\;b_\alpha \;\;
\graphical[scale=1.25]{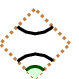}
\end{equation*}
where \(\alpha\) is the spectral parameter with a normalization yet to 
be fixed. The axis connects two opposite corners of the plaquette, one
of which is marked with a tag to track the orientation of the 
plaquette. For a rhombic embedding of the covering graph on
the plane, we call the angle at this corner the opening angle of the 
plaquette. For the model to be integrable, the weights must satisfy the 
Yang-Baxter equation
\begin{equation}
\label{eq:YangBaxterEquation}
\graphical[angle=90]{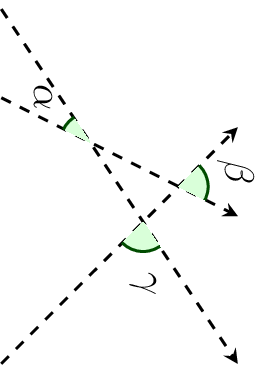} = \graphical[angle=90]{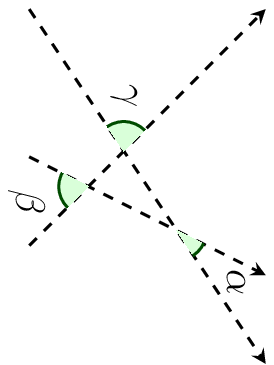}
\end{equation}
as well as the inversion relation,
\begin{equation}
\label{eq:InversionRelation}
\graphical{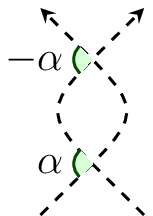} = %a_\alpha a_{-\alpha} 
\graphical{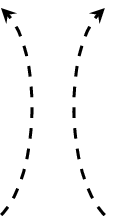}
\end{equation}

\begin{figure}
\begin{equation*}
\begin{array}{ccccccccc}
\graphical{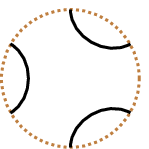} &\quad& \graphical{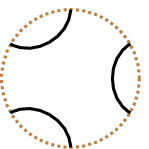} &\quad& 
\graphical{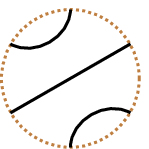} &\quad& 
\graphical{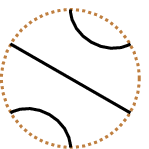} &\quad& \graphical{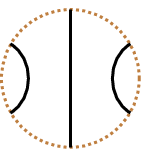}  \\
\mathrm{I} && \mathrm{II} && \mathrm{III} && \mathrm{IV} && \mathrm{V}
\end{array}
\end{equation*}
\caption{The 5 different chord diagrams between 6 points that form
the basis in the formal expansion of the Yang-Baxter equation.}
\label{fig:chords}
\end{figure}

The Temperley-Lieb algebra \cite{TL-RBPC} allows formal expansion
of the diagrams in the 5 possible connectivities, also 
known as chord diagrams, that correspond to the different pairings
between 6 points (\fref{fig:chords}). 
In general, the number of pairings $(2j)!/(j+1)!j!$ of $2j$ points on a 
circle by non-intersecting chords is known as the $j^{\mathrm{th}}$ 
Catalan number in combinatorics.
 
The resulting relations are
\begin{eqnarray}
a_\alpha\;a_\beta\;a_\gamma=a_\alpha\;b_\beta\;b_\gamma
+b_\alpha\;a_\beta\;b_\gamma+b_\alpha\;b_\beta\;a_\gamma 
+ n\;b_\alpha\;b_\beta\;b_\gamma\label{eq:PottsYangBaxter}\\
a_\alpha\;b_{-\alpha} + b_\alpha\;a_{-\alpha} + n\;a_\alpha\;a_{-\alpha} 
= 0\label{eq:PottsInversionRelation}
\end{eqnarray}
respectively, where the normalization of the spectral parameters is 
\begin{equation}
\label{eq:normalization}
\alpha + \beta + \gamma = 2\pi
\end{equation}
suggesting a possible interpretation as angles.

%%%%%%%%%%%%%%%%%%%%%%%%%%%%%%%%%%%%%%%%%%
%%%%%%%%%%%%%%%%%%%%%%%%%%%%%%%%%%%%%%%%%%

\subsection{The loop observable}

%%%%%%%%%%%%%%%%%%%%%%%%%%%%%%%%%%%%%%%%%%
\begin{figure}
\begin{center}
\includegraphics{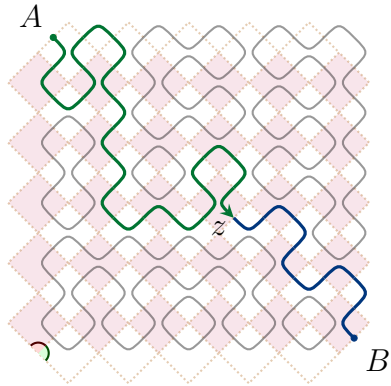}
\end{center}
\caption{Contribution to the observable $\psi(z)$ from a loop 
configuration on a regular square lattice. The thick curve is the
exploration path that runs from $A$ to $B$ and has $z$ lying on it.
The shaded plaquettes
correspond to the vertical, and the white plaquettes to the horizontal
edges in the original Potts spin model. The marked angles at the
lower left corner show that the opening angles of the two classes
or rhombi are supplimentary. The winding angle is the total angle that
the curve turns to get from $A$ to $z$, which, in this configuration, 
is zero.}
\label{fig:denseobservable}
\end{figure}
%%%%%%%%%%%%%%%%%%%%%%%%%%%%%%%%%%%%%%%%%%

A discretely holomorphic observable 
\cite{RC-HPPM,IC-DHPILM} of the
model, for a rhombic embedding of the covering lattice on
the complex plane,  is defined on the midpoint of the sides of the 
plaquettes,
\[
\psi(z) = \sum_{\gamma\;\in\;\Gamma(z)}\ee^{-\ii\;\sigma
\;W_\gamma(A,\,z)
}\;w(\gamma)
\]
Here $\sigma$ is a real constant called the
conformal spin of the observable, $\Gamma(z)$ is the set of
all the configurations having a curve, called the exploration path,
between two fixed points $A$ and $B$ on the boundary with 
$z$ lying on it, $w(\gamma)$ is the weight
of the configuration $\gamma$, and $W_\gamma(A,z)$ is the winding angle
of the exploration path going from $A$ to $z$ 
(\fref{fig:denseobservable}). That is, on each plaquette it obeys
\[
\sum_{\lozenge}\psi(z)\;\Delta z = 0
\]
where the $\Delta z$ are the differences in complex coordinates of the
endpoints of the side $z$ is on, traversed in the anticlockwise 
direction around the rhombus $\lozenge$. 
Consequently such a sum around any closed path on the
covering lattice vanishes to give a discrete analogue of Cauchy's 
integral theorem for holomorphic functions.

\begin{figure}
\begin{center}
\begin{tabular}{ccc}
$\graphical{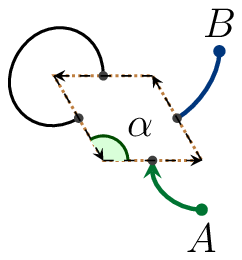}$ & $\quad\quad$ & $\graphical{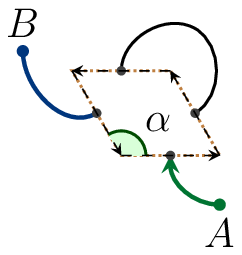}$
\end{tabular}
\end{center}
\caption{The external configurations for equations
\eref{eq:one} and \eref{eq:two}
respectively.  
The discrete contour consisting of the $\Delta z$ is indicated by 
dashed arrows.}
\label{fig:denseonerhombus}
\end{figure}

To establish the claim, Ikhlef and Cardy \cite{IC-DHPILM} 
considered the contour sum around a single rhombic
plaquette with opening angle $\alpha$, breaking up the configuration
into two parts for the interior and the exterior of the 
rhombus respectively, and grouping the terms by the exterior 
configuration, which thus for the purposes of the argument can be 
thought of as frozen henceforth, and then by the point of first entry 
into the rhombus for the curve. Therefore it suffices for the sum to 
vanish for each of these groupings individually. Defining an auxiliary
function
\[
\phi(\alpha) = \ee^{\ii\;(1-\sigma)\;\alpha}
\]
for notational convenience, we have, for the two possible connectivities 
of the external 
configurations (\fref{fig:denseonerhombus}), 
\begin{eqnarray}
\label{eq:one}n\;\wrap{1 - \phi(\alpha-\pi)}\;a_\alpha
 + \wrap{1 - \phi(\alpha-\pi) + \phi(-\pi) - \phi(\alpha)}\;b_\alpha 
 = 0\\
\wrap{1-\phi(\alpha-\pi) + \phi(\pi) - \phi(\alpha)}\;a_\alpha 
+ n\;\wrap{1-\phi(\alpha)}\;b_\alpha = 0\label{eq:two}
\end{eqnarray}
For non-trivial solutions to exist, the determinant of the system
must vanish, and we take the factors of the determinant independent
of $\alpha$ to be zero, so that the condition is satisfied on all
rhombuses, to obtain 
\[
n^2 - 2 = \phi(\pi) + \phi(-\pi)
\]
and therefore,
\[
1-\sigma = \frac{2\lambda}{\pi}-2\ell\quad\quad(\ell \in \mathbb{Z})
\]
which results in a family of solutions labelled by an integer $\ell$,
\[
\frac{a_\alpha}{b_\alpha} = (-1)^{\ell} \, 
\frac{\sin\wrap{(\frac{\lambda}{\pi} - \ell)\alpha}}
{\sin\wrap{(\frac{\lambda}{\pi} - \ell)\,(\pi-\alpha)}}
\]
all of which satisfy the integrability condition as shown below.
The standard solution $a = \sin u$, $b = \sin(\lambda - u)$ 
corresponds to $\ell = 0$, and lets us identify the angle,
in this case, as a 
(re-scaled) spectral parameter,
\[
u = \frac{\lambda}{\pi}\,\alpha
\]
It was observed long ago \cite{KP-SDCA} that this is
the required embedding of the anisotropic lattice for correlations
to be conformally, or more simply, rotationally, invariant.
The criticality condition
\[
\frac{a_{\pi-\alpha}}{b_{\pi-\alpha}}\,\frac{a_\alpha}{b_\alpha} = 1
\]
can be immediately verified, and enables the identification
of these supplementary weights to be the dual weights in the other 
lattice direction (\fref{fig:denseobservable}). Thus the role of 
the crossing parameter is played by $\pi$ in this normalization.

%%%%%%%%%%%%%%%%%%%%%%%%%%%%%%%%%%%%%%
%%%%%%%%%%%%%%%%%%%%%%%%%%%%%%%%%%%%%%

It is instructive to compare, once a suitable observable has been
identified, the problem of solving the discrete holomorphicity condition
which is linear in the weights with a given spectral parameter, 
with the usual one of finding a solution of the 
Yang-Baxter equation which is cubic in the weights and relates weights 
with different spectral parameters.

\subsection{Rhombic embeddings of Baxter lattices}
\label{sec:embedding}

The appearance of rhombic embeddings, or equivalently,
isoradial embeddings, in the above construction is
doubly propitious, as on one hand, discrete complex analysis 
on them is natural \cite{D-PTRL}, 
and on the other hand, the key class of lattices for integrable
models \cite{B-S8V}, usually called the Baxter lattices, admits such 
embeddings \cite{KS-REPQG}. Our argument draws from related ideas that 
have appeared in geometrical characterization of Baxter lattices
\cite{CS-ZIIM}, and emergence of discrete 
conformal symmetry in solutions of the star-triangle relation 
\cite{BSM-FVYB}.

%%%%%%%%%%%%%%%%%%%%%%%%%%%%%%%%%%%%%%%%%
\begin{figure}
\begin{center}
 \includegraphics{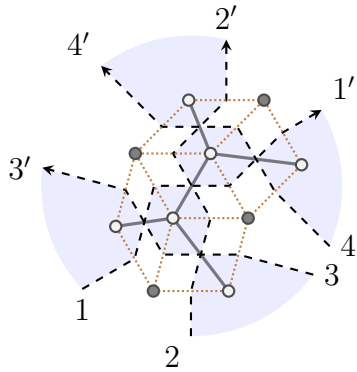}
\end{center}
\caption{Part of a Baxter lattice with its rhombic embedding. The
rapidity lines are dashed. One of the bipartite classes of regions 
is shaded. The spins are on the white circles, and the disorder 
variables are on the shaded circles. The thick lines are edges 
indicating interaction between spins. The dotted lines show the 
associated rhombic embedding with each rhombus having an interaction
edge as a diagonal that we call its axis. }
\label{fig:embedding}
\end{figure}
%%%%%%%%%%%%%%%%%%%%%%%%%%%%%%%%%%%%%%%%%

We start, therefore, with a Baxter lattice (\fref{fig:embedding}). It 
is a planar graph defined by a set of directed lines called rapidity 
lines, which are usually straight, but for our purposes it is only 
necessary to keep the topological characterization: no line crosses 
itself or is periodic, and no two lines cross each other more than once. 
The lines divide the plane into bipartite regions, and for spin models 
such as the Potts model, we place spins, which are the 
nodes of the graph, on one class of regions, and disorder variables, 
which are the faces of the graph, on the other class. At each crossing
of rapidity lines, where four regions of alternating classes meet,
an edge connects the two spins to indicate interaction between them.
If we place a parameter, called the rapidity parameter, on each 
line, then the interaction on an edge can depend on the difference, 
called the spectral parameter, of the parameters of the two crossing 
lines.

The covering lattice of this graph is a quad-graph whose nodes are 
the spins and the disorder variables, and these are connected
by an edge if the two containing regions are adjacent, so each edge
connects a spin with a disorder variable. Each face then has as one
diagonal an edge of the original lattice which we call the axis.
In the terminology of \cite{KS-REPQG}, a train-track is a sequence
of adjacent faces of a quad-graph which upon entering a face via an edge
exits through the opposite edge, and a quad-graph has a rhombic
embedding if and only no train-track crosses itself or is periodic,
and no two train-tracks cross each other more than once. The 
identification of train-tracks with rapidity lines is thus immediate,
and since in a rhombic embedding all the edges that cross a rapidity
line are parallel, we can identify (the complex exponential of
the imaginary unit times) the rapidity parameter with the
common direction, called the transversal, of the edges, treated
as a unimodular complex number. The spectral 
parameter is then naturally given by the angle between the two 
transversals of a rhombus, or its supplement, 
depending on the direction of crossing and the placement of the axis.
Since the common length of the sides of all the rhombus 
provide the only length scale of the problem, we set it to unity without
loss of generality.

The conditions of integrability, that is, the Yang-Baxter equation 
\eref{eq:YangBaxterEquation}
and the inversion relation \eref{eq:InversionRelation}, ensure that 
the partition function of a Baxter lattice remains unchanged, 
or possibly acquires a trivial 
constant factor, as the rapidity lines are moved around topologically,
that is, by the third and the second Reidemeister moves respectively, 
while the end points of the lines are kept fixed. 
This property is referred to
as $Z$-invariance in the literature. In other words,
the partition function is a function only of the permutation (denoted
by primes in \fref{fig:embedding}), or more generally, the braiding, of
the rapidity lines, and is
independent of the internal lattice. Our identification of rapidity
parameters and transversals implies that the boundary of the rhombic
domain remains unchanged by such moves, and the boundary can thus be
regarded as belonging to the equivalence class of such lattices.

\subsection{Derivation of integrability}

The proof of discrete holomorphicity of the observable by grouping 
terms by external connectivities suggests a natural extension to any
simply-connected domain of rhombic faces. A simple proof is given by
induction on the number of faces, and proceeds by dividing the domain
into smaller pieces, and regarding all but one of the pieces as part of
the external configuration. 

%%%%%%%%%%%%%%%%%%%%%%%%%%%%%%%%%%%%%%%%%
\begin{figure}
\begin{equation*}
\fl
\graphical{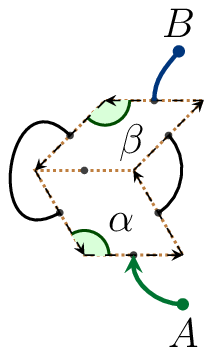} = a_{\alpha}\;\graphical{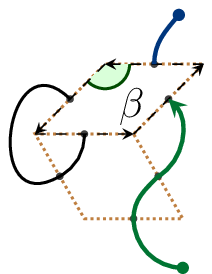}
+ b_{\alpha}\;\graphical{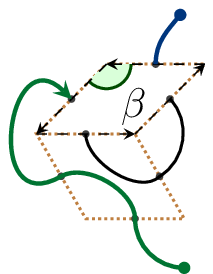} + a_{\beta}\;\graphical{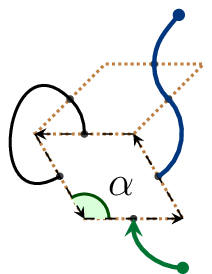}
+ b_{\beta}\;\graphical{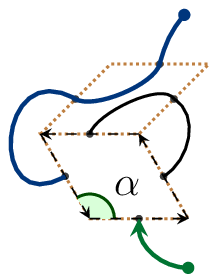}
\end{equation*}
\caption{The contour sum around two rhombuses. The dashed arrows
show the discrete contours. The labels on the endpoints of the curve
on the right hand side has been omitted for clarity.}
\label{fig:tworhombi}
\end{figure}
%%%%%%%%%%%%%%%%%%%%%%%%%%%%%%%%%%%%%%%%%

To illustrate the procedure, we take the domain to consist of two 
rhombuses (\fref{fig:tworhombi}), having opening angles $\alpha$ and
$\beta$, with an external connectivity of 
chord diagram V (\fref{fig:chords}). The figure denotes the discrete 
contour sum of the observable  around the shown contour, which 
can be broken up into contour sums around each rhombus. We then expand 
the external rhombus into its connectivities. Evidently, the total sum 
reduces to a linear combination of sums around only one rhombus, all of 
which are known to vanish, regardless of the entry points. Therefore, 
the total sum itself vanishes for the given external configuration. 
We thus conclude that the discrete contour sum of the observable around 
any simply-connected domain vanishes for each of the external 
connectivities individually.

Evaluating the left hand side of \fref{fig:tworhombi} directly, 
that is, setting the contour sum of the observable 
around two rhombuses with the external chord diagram V to zero, 
we arrive at
\begin{eqnarray*}
\fl
\left(\phi(\pi-\beta)-\phi(\alpha-\pi)\right)\,n^2\,a_\alpha\,a_\beta
+\left(\phi(-\beta)-\phi(\alpha)\right)\,n^2\,b_\alpha\,b_\beta\\
+\left(\phi(\pi-\beta)+\phi(-\beta)-\phi(\alpha)-\phi(\alpha-\pi)\right)
\wrap{\,n\,a_\alpha\,b_\beta
+n\,b_\alpha\,a_\beta}
&=0
\end{eqnarray*}
which is quadratic in the weights. Putting $\beta = -\alpha$ gives 
\begin{eqnarray*}
\left(\phi(\pi+\alpha)-\phi(\alpha-\pi)\right)\,n\,
\left(
n\,a_\alpha\,a_{-\alpha}+a_\alpha\,b_{-\alpha}+b_\alpha\,a_{-\alpha}
\right)
&=0
\end{eqnarray*}
from which we immediately conclude that 
\begin{equation*}
a_\alpha\,b_{-\alpha}+b_\alpha\,a_{-\alpha}+n\,a_\alpha\,a_{-\alpha} = 0
\end{equation*}
that is, the inversion relation \eref{eq:InversionRelation} holds.

%%%%%%%%%%%%%%%%%%%%%%%%%%%%%%%%%%%%%%%%%
\begin{figure}
\begin{center}
\includegraphics{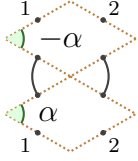}
\end{center}
\caption{A pair of ghost rhombuses. The internal unnamed points 
connected by the thick lines are indentified. The external numbered
points on the plane are also identified, allowing the pair to 
be folded up.}
\label{fig:ghost}
\end{figure}
%%%%%%%%%%%%%%%%%%%%%%%%%%%%%%%%%%%%%%%%%

It is interesting to note that the inversion relation is at odds with
the premises of the rhombic embedding theorem in that the two 
rapidity lines cross twice, and hence, the relation does not admit an
embedding, in contrast to the Yang-Baxter equation below. 
However, the appearance of the negative angle, or 
equivalently, the negative area, may be interpretated as the result
of flipping a plaquette over to reveal another from underneath
(\fref{fig:ghost}), whose
contributions to the weight of the configuration cancel. Also,
with this interpretation, the four external midpoints are, in fact,
superposed to give just two midpoints on adjacent sides, perhaps 
on some other rhombus, and therefore the relation permits the 
insertion of a pair of rhombuses with opposite angles at will, 
as in the standard proof of commutating transfer matrices. We also
note that such topological modifications to the Baxter lattice
respect the $Z$-invariance, or the invariance of the 
partition function, and can serve to broaden the equivalence class
of lattices having the same embedding, once such ghost pairs of 
rhombuses are allowed.

\begin{figure}
\begin{center}
\includegraphics{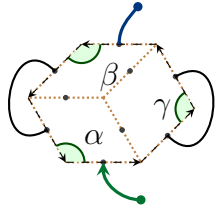}
\end{center}
\caption{The hexagon formed by three rhombuses with opening angles
that sum up to $2\pi$. The contour sum is around the rhombus with a 
particular external chord diagram that produces the Yang-Baxter equation
directly.}
\label{fig:cheating}
\end{figure}

To prove the Yang-Baxter relation, we consider a third rhombus with
opening angle $\gamma$ to complete the hexagon, resulting in the
geometric requirement, equation \eref{eq:normalization}, that relates
the spectral parameters. Again, evaluating the contour sum of the 
observable around the hexagon (\fref{fig:cheating}) for external chord 
diagram V, we get
\[
\left(\phi(-\beta)-\phi(\alpha)\right)\,n^2\,
\mathrm{YB}(\alpha, \beta, \gamma)=0
\]
where we have defined
\[
 \mathrm{YB}(\alpha, \beta, \gamma) := b_\alpha\,b_\beta\,a_\gamma
 +b_\alpha\,a_\beta\,b_\gamma
+a_\alpha\,b_\beta\,b_\gamma+n\,b_\alpha\,b_\beta\,b_\gamma
-\,a_\alpha\,a_\beta\,a_\gamma
\]
with the understanding that the angles follow equation 
\eref{eq:normalization} so that there are really two independent 
variables. Therefore we immediately conclude that
\[
 \mathrm{YB}(\alpha, \beta, \gamma) = 0
\]
or in other words, equation \eref{eq:PottsYangBaxter} holds. Here,
and in what follows, including the appendix, we avoid special treatment 
of isolated values of $n$ for which the proofs can be adopted, or an 
appeal to continuity can be made.

\begin{figure}
\begin{center}
$\graphical{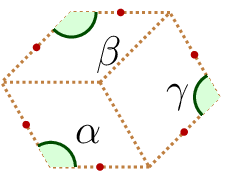} \quad=\quad 
\graphical{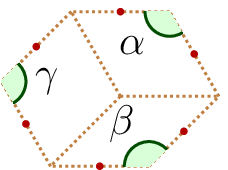}$
\end{center}
\caption{The two ways to tile the hexagon with three rhombuses. 
For the Potts spin model, the left arrangement makes a star, and the
right makes a triangle of interaction edges, and so we refer to the two
arrangements as such. The dots mark the observables on the boundary.}
\label{fig:startriangle}
\end{figure}

However, in light of the result to follow in \sref{sec:invariance}, we 
have an
alternate proof by considering the two different arrangements of 
the rhombuses to produce the same hexagon (\fref{fig:startriangle}).
For each of the external connectivities (\fref{fig:chords}), we take the 
contour sum of the two sides, and take the difference.
Since both of the sums vanish, so does the difference, and thus we have 
\begin{eqnarray*}
 \left(n\,\phi (\alpha -2 \pi )
 -n\,\phi (\alpha )+n^3 \,\phi (-\beta )-n^3\right)\, 
 \mathrm{YB}(\alpha ,\beta ,\gamma ) = 0\\
n^2 (\phi (-\beta )-\phi (2 \pi -\beta )) 
\,\mathrm{YB}(\alpha ,\beta ,\gamma ) = 0\\
n^2 (\phi (\alpha -2 \pi )-\phi (\alpha ))\, 
\mathrm{YB}(\alpha ,\beta ,\gamma ) = 0\\
 \left(n^3-n\,\phi (2 \pi -\beta )
 +n\,\phi (-\beta )-n^3 \phi (\alpha )\right)\, 
 \mathrm{YB}(\alpha ,\beta ,\gamma )=0\\
2 n^2 (\phi (-\beta )-\phi (\alpha )) 
\,\mathrm{YB}(\alpha ,\beta ,\gamma ) = 0
\end{eqnarray*}
for all $\alpha$ and $\beta$, and therefore all of them show that
the Yang-Baxter equation, or the star-triangle relation, holds.

%%%%%%%%%%%%%%%%%%%%%%%%%%%%%%%%%%%%%%%%%
%%%%%%%%%%%%%%%%%%%%%%%%%%%%%%%%%%%%%%%%%

%%%%%%%%%%%%%%%%%%%%%%%%%%%%%%%%%%%%%%%%%
%%%%%%%%%%%%%%%%%%%%%%%%%%%%%%%%%%%%%%%%%

\section{The dilute loop model}\label{sec:dilute}
\subsection{Yang-Baxter integrability}
\label{sec:DiluteYangBaxter}
The dilute, or $O(n)$, loop model arises as the high-temperature
expansion of an $n$-component vector spin model on the plane that 
is invariant under the action of the rotation group. 
In this model, each closed loop has a fugacity, or weight 
\[
n = -2\cos4\eta \quad\quad 
(-2\le n \le 2, \quad 0\le\eta\le\frac{\pi}{4})
\]
and the weights of the configurations on the plaquettes are encoded
in the $R$-matrix as a linear combination of the 9 connectivities 
between 4 points, where the non-intersecting curves are allowed to be
absent, with
\begin{equation*}
\fl
\graphical{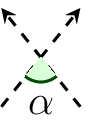} = t_{\alpha}\,\graphical{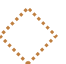} 
+ {u_1}_{\alpha}\,\wrap{\graphical{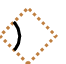} + \graphical{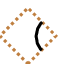}}
+ {u_2}_{\alpha}\,\wrap{\graphical{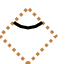} + \graphical{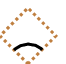}}
+ v_{\alpha}\,\wrap{\graphical{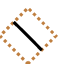} + \graphical{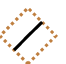}}
+ a_{\alpha}\,\graphical{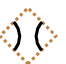}
+ b_{\alpha}\,\graphical{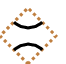}
\end{equation*}
The configurations where a curve starts or ends abruptly in the bulk
are not allowed and can be assigned a null weight. The generality
sacrificed in favour of brevity in the choice of the weights is
ultimately justified by the obtained solutions.

Expanding the Yang-Baxter equation \eref{eq:YangBaxterEquation} in 
this $R$-matrix, we get 51 inequivalent connectivities
between 6 points and their corresponding equations, which reduce to
\begin{eqnarray*}
\fl
\mathrm{YB}_1(\alpha,\beta,\gamma)\,:=
\,&{u_2}_{\alpha}\,{u_2}_{\beta}\,a_{\gamma}
+n\,{u_2}_{\alpha}\,{u_2}_{\beta}\,b_{\gamma}
+t_{\alpha}\,t_{\beta}\,{u_2}_{\gamma}
-{u_1}_{\alpha}\,{u_1}_{\beta}\,t_{\gamma}
-v_{\alpha}\,v_{\beta}\,{u_2}_{\gamma} = 0\\
\fl\mathrm{YB}_2(\alpha,\beta,\gamma)\,:=
\,&t_{\alpha}\,{u_1}_{\beta}\,v_{\gamma}
+{u_2}_{\alpha}\,v_{\beta}\,{u_1}_{\gamma}
-t_{\gamma}\,{u_1}_{\beta}\,v_{\alpha}
-{u_1}_{\alpha}\,v_{\beta}\,{u_2}_{\gamma} = 0\\
\fl\mathrm{YB}_3(\alpha,\beta,\gamma)\,:=
\,&{u_2}_{\alpha}\,b_{\beta}\,a_{\gamma}
+{u_2}_{\alpha}\,a_{\beta}\,b_{\gamma}
+n\,{u_2}_{\alpha}\,b_{\beta}\,b_{\gamma}
+t_{\alpha}\,{u_2}_{\beta}\,{u_2}_{\gamma}
-a_{\alpha}\,{u_1}_{\beta}\,{u_1}_{\gamma} = 0\\
\fl\mathrm{YB}_4(\alpha,\beta,\gamma)\,:=
\,&v_{\alpha}\,{u_1}_{\beta}\,b_{\gamma}
+{u_1}_{\alpha}\,v_{\beta}\,{u_2}_{\gamma}
-a_{\alpha}\,{u_1}_{\beta}\,v_{\gamma} = 0\\
\fl\mathrm{YB}_5(\alpha,\beta,\gamma)\,:=
\,&{u_1}_{\alpha}\,b_{\beta}\,{u_1}_{\gamma}
+v_{\alpha}\,{u_2}_{\beta}\,v_{\gamma}
-a_{\alpha}\,{u_2}_{\beta}\,a_{\gamma} = 0\\
\fl\mathrm{YB}_6(\alpha,\beta,\gamma)\,:=
\,&b_{\alpha}\,b_{\beta}\,a_{\gamma}
+b_{\alpha}\,a_{\beta}\,b_{\gamma}
+a_{\alpha}\,b_{\beta}\,b_{\gamma}
+n\,b_{\alpha}\,b_{\beta}\,b_{\gamma}\,
+{u_2}_{\alpha}{u_2}_{\beta}\,{u_2}_{\gamma}
-a_{\alpha}\,a_{\beta}\,a_{\gamma} = 0
\end{eqnarray*}
subject to equation \eref{eq:normalization}, upon permuting $\alpha$, 
$\beta$ and $\gamma$. We will refer to the functions defined, which we
need to show to vanish, as the $\mathrm{YB}_i(\alpha,\beta,\gamma)$ 
functions. In general, 
the number of ways of drawing non-intersecting chords on a circle 
between $j$ points, 
$\sum_{k=0}^{\lfloor j/2\rfloor }{{j}\choose{2k}}(2j)!/(j+1)!j!$, 
is known as the $j^{\mathrm{th}}$ Motzkin number in combinatorics.
Here 51 is the $6^{\mathrm{th}}$ Motzkin number.

\subsection{The loop observable}

\begin{figure}
\begin{center}
\includegraphics{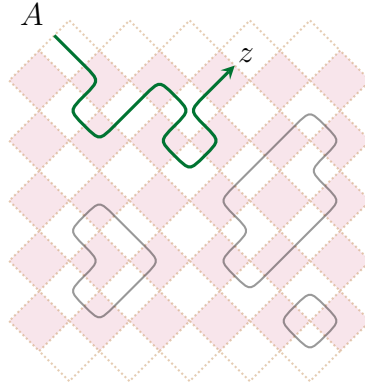}
\end{center}
\caption{Contribution to the loop observable $\psi(z)$ for a 
configuration $\gamma$ on a regular square lattice. The winding angle
$W_{\gamma}(A, z)$ is the total angle the exploration path turns to 
get from the
boundary point $A$ to the specified point $z$, which in this figure 
is $\pi/2$.}
\label{fig:diluteobservable}
\end{figure}

The holomorphic observable \cite{IC-DHPILM} of the model is defined
on the midpoints of the sides of the plaquettes to be 
\[
\psi(z) = \sum_{\gamma\;\in\;\Gamma(z)}\ee^{-\ii\;\sigma
\;W_\gamma(A,\,z)
}\;w(\gamma)
\]
where $\Gamma(z)$ now is the set of configurations
that have an open path from $A$ to $z$, which is
the exploration path for this case, and the rest
of the symbols are analogous to those in the dense loop model. Again,
we want 
\[
\sum_{\lozenge}\psi(z)\;\Delta z = 0
\]
on a rhombus of opening angle $\alpha$, and to this end, we consider
grouping the terms by the specific point through which the path
enters the plaquette. With one point already accounted for, the
remaining 3 points can have 4 connectivites (the $3^{\mathrm{rd}}$
Motzkin number),
and consequently, we have the discrete holomorphicity relations 
as in \cite{IC-DHPILM},
\begin{eqnarray}
\label{eq:diluteone}
{t}_{\alpha}-\phi(\alpha-\pi)\,{u_1}_{\alpha}
-\phi(\alpha)\,{u_2}_{\alpha}-{v}_{\alpha}=0\\
\label{eq:dilutetwo}
\phi(\pi)\,{u_1}_{\alpha}+n\,{u_2}_{\alpha}
+\phi(\alpha-2\pi)\,{v}_{\alpha}-\phi(\alpha)\,{a}_{\alpha}
-n\,\phi(\alpha)\,{b}_{\alpha}=0\\
\label{eq:dilutethree}
\phi(\alpha+\pi)\,{u_1}_{\alpha}+\phi(\alpha-2\pi)\,{u_2}_{\alpha}
+n\,{v}_{\alpha}-\phi(2\pi)\,{a}_{\alpha}-\phi(-2\pi)\,{b}_{\alpha}=0\\
\label{eq:dilutefour}
n\,{u_1}_{\alpha}+\phi(-\pi)\,{u_2}_{\alpha}
+\phi(\alpha+\pi)\,{v}_{\alpha}
-n\,\phi(\alpha-\pi)\,{a}_{\alpha}-\phi(\alpha-\pi)\,{b}_{\alpha}=0
\end{eqnarray}
with the same definition of $\phi(\alpha)$ as before. Considering
the other points the curve could enter through, we find that the 
complex conjugates of the equations are also satisfied, which ensures
that the weights are real. Consequently we have 8 real equations for
6 real variables, but fortunately 2 of them are superfluous, and the
system has non-trivial solutions when the determinant of the reduced
system vanishes, a condition that we ensure by choosing the $\alpha$ 
independent factor to be zero, which results in
\[
\phi(4\pi)+\phi(-4\pi) = 3 n - n^3
\]
so that
\[
1 - \sigma = \frac{3\eta}{\pi} + \frac{1}{2}\ell
\quad\quad(\ell \in \mathbb{Z})
\]

%%%%%%%%%%%%%%%%%%%%%%%%%%%%%%%%%%%%%%%%
\begin{figure}
\begin{equation*}
\begin{array}{ccccccccc}
\includegraphics{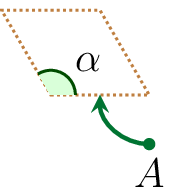} &\quad& 
\includegraphics{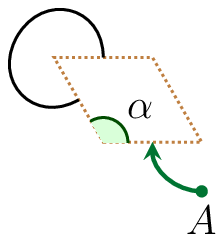}
& \quad&\includegraphics{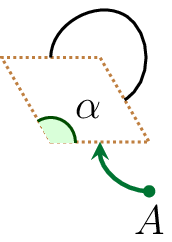} & \quad&
\includegraphics{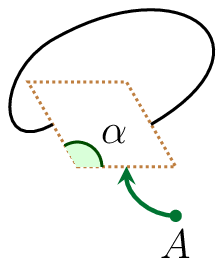}
\end{array}
\end{equation*}
\caption{The external configurations for  equations
\eref{eq:diluteone}, \eref{eq:dilutetwo}, \eref{eq:dilutethree},
and \eref{eq:dilutefour} respectively.
The embedding of the external curves must be so that they do not
enclose the point $A$ on the boundary, which uniquely specifies the
winding angles.}
\label{fig:diluteonerhombus}
\end{figure}
%%%%%%%%%%%%%%%%%%%%%%%%%%%%%%%%%%%%%%%%

Thus we again find a series of solutions to the integrability 
conditions labelled by an integer $\ell$, as is proved below. However,
for clarity, we restrict ourselves in this section to the $\ell = 0$ 
case, for which the solution is 
\begin{eqnarray*}
t_\alpha
=\sin\frac{3\eta}{\pi}\alpha\,\sin\frac{3\eta}{\pi}(\pi-\alpha)
+\sin2\eta\,\sin3\eta\\
{u_1}_\alpha=\sin\frac{3\eta}{\pi}\alpha\,\sin2\eta\\
{u_2}_\alpha=\sin\frac{3\eta}{\pi}(\pi-\alpha)\,\sin2\eta\\
v_\alpha
=\sin\frac{3\eta}{\pi}\alpha\,\sin\frac{3\eta}{\pi}(\pi-\alpha)\\
a_\alpha
=\sin\frac{3\eta}{\pi}\alpha
\,\sin\left(\frac{3\eta}{\pi}\alpha-\eta\right)\\
b_\alpha
=\sin\frac{3\eta}{\pi}(\pi-\alpha)
\,\sin\left(\frac{3\eta}{\pi}(\pi-\alpha)-\eta\right)
\end{eqnarray*}
which is indeed the standard solution \cite{N-CMOM}, with 
\[
u = \frac{3\eta}{\pi}\alpha  
\]
Note that the convexity requirement on the 
rhombuses, $0 < \alpha < \pi$, guarantees positivity of the weights
for the $\ell = 0$ solution, for both the dilute and the dense loop 
models. Also, it is reasonable to expect that with a square
lattice with anisotropic interaction, the correlation functions will
become rotationally invariant when it is embedded on the plane according
to this prescription for the angles.

It is interesting to note that the existence of discretely 
holomorphic observables indicate an explanation of the fact that
solvable critical weights are usually trigonometric. Since the
real weights have to obey homogeneous linear relations with
coefficients that are sums of pure complex phases, their solutions
are real-valued rational functions of those phases, and thus
trigonometric functions are the most natural candidates. This
provides a strong motivation to look for such observables in 
critical solvable models which are known to have trigonometric weights,
especially those with their continuum limits described by minimal
models.

\subsection{Derivation of integrability}
The extension of the vanishing contour sum property to
simply-connected domains with specified external configuration 
applies \textit{mutatis mutandis} to the dilute model. We therefore
consider \fref{fig:startriangle} in this context where the curve
enters via a specified point, say, the base. The remaining 5 points
may have 21 connectivities (the $5^{\mathrm{th}}$ Motkzkin number), 
and the resulting
equations, which are the differences of the contour sums around the
hexagon for the two arrangements, are omitted for brevity. However,
in these equations, the weights enter only through the
$\mathrm{YB}_i(\alpha, \beta, \gamma)$ functions. They are therefore
an over-determined linear system in the functions that in fact, 
as shown in the Appendix, forces them to vanish, proving the
Yang-Baxter relations.

\section{Invariance of the observable at the boundary}
\label{sec:invariance}

%%%%%%%%%%%%%%%%%%%%%%%%%%%%%%%%%%%%%%%%
\begin{figure}
\begin{center}
\begin{tabular}{ccccc}
\includegraphics{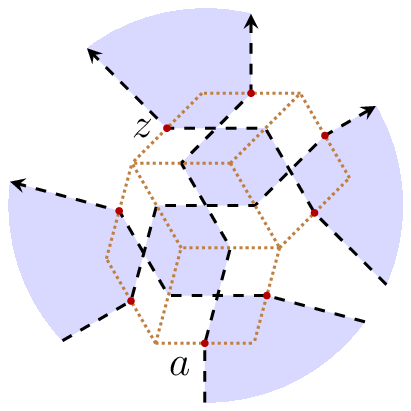} & & 
\includegraphics{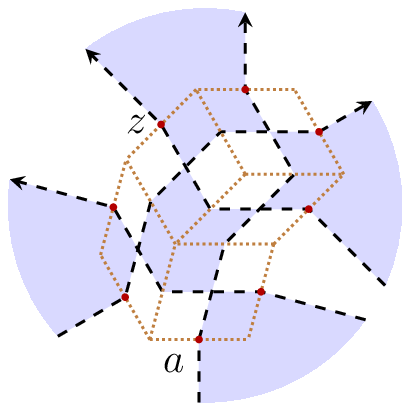} & &
\includegraphics{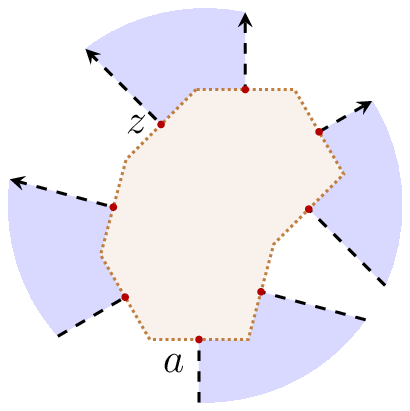} \\
$\Omega_1$ & & $\Omega_2$ & & $\partial \Omega$
\end{tabular}
\end{center}
\caption{Two equivalent Baxter sublattices, $\Omega_1$ and $\Omega_2$,
and their common boundary $\partial \Omega$. The highlighted points
are the midpoint of the sides of the rhombuses that are on 
$\partial \Omega$. The exploration path first enters 
the sublattice via the boundary point $a$. The value of the observable 
at a boundary point $z$ remains the same for the two lattices
and unambiguously defines it on $\partial \Omega$ without referring
to the interior.}
\label{fig:invariance}
\end{figure}
%%%%%%%%%%%%%%%%%%%%%%%%%%%%%%%%%%%%%%%%

Once the integrability conditions are derived from the holomorphicity
condition, however, an alternative explanation of the proof can be
deduced.  Inside a Baxter lattice, we consider a sublattice $\Omega_1$, 
and modify it by Reidemeister moves of the 
rapidity lines that leave the exterior of the sublattice unaltered
(\fref{fig:invariance}), to obtain another sublattice $\Omega_2$. With 
our identication of rapidity parameters with transversals in 
\sref{sec:embedding}, the boundary of the sublattices in a rhombic
embedding remains the same, and we denote it by $\partial \Omega$.
The effect of the modifications on the embedding is to shuffle rhombuses
around, as is evident from \fref{fig:startriangle}.

Now compare, for the two sublattices, the observable at a midpoint $z$ 
on $\partial \Omega$, 
\begin{eqnarray*}
\psi_{\Omega}(z) = \sum_{a\in \partial \Omega} \Bigg(
\sum_{\gamma\;\in\;\Gamma_{\Omega}(z; a)}
\ee^{-\ii\;\sigma\;W_\gamma(A,\,z)}\;w(\gamma)\Bigg)
\end{eqnarray*}
where $\Omega$ stands for either of the two sublattices, and
$\Gamma_{\Omega}(z; a)$ is the subset of $\Gamma_{\Omega}(z)$ for which 
$a$ is the point of entry of the exploration path into the interior.
Clearly, it suffices to show that the quantity in parentheses is equal
for the two sublattices to prove that 
\[
\psi_{\Omega_1}(z) = \psi_{\Omega_2}(z)
\]
and consequently, for every member of the equivalence class of the
sublattices, and hence, $\psi$ is unambiguously defined on 
$\partial \Omega$, independent of its interior lattice structure.

The exterior common to the two sublattices can be factored out by 
grouping the terms by the configurations on it, which fixes the value of
the winding angle up to $a$, and thus can be factored out. The only 
information we need about this configuration is then its connectivity
given by a chord diagram appropriate for the model, which we refer to as 
the external chord diagram. Note that for the dilute model, the diagrams
that connect $a$ to another point on $\partial \Omega$ are not allowed 
as external diagrams.

Hence we are left with comparing, for the two sublattices, 
\[
\sum_{\gamma\;\in\;\Gamma_{\Omega}^{\mathrm{int}}(z; a)}
\ee^{-\ii\;\sigma\;W_\gamma(a,\,z)}\;w(\gamma)
\]
where a given external diagram is imposed. Here  
$\Gamma_{\Omega}^{\mathrm{int}}(z; a)$ is the set of interior
configurations on the sublattice $\Omega$ where $z$ and $a$ lie on the
same curve, possibly going through both the interior and the exterior
of the sublattice. This time we group the terms in the sum by the 
internal chord diagrams, that is, the connectivity between the points
on $\partial \Omega$ in the internal configurations, for the comparison.
For each group, the winding angles are the same because the curve follow
identical chords on both of the chord diagrams which, by the existence 
of the common boundary $\partial \Omega$, have well-defined winding 
angles for each segment of the curve joining two points on it.

The quantity left to compare is the sums of weights of all 
configurations with the same chord diagrams for the two sublattices. 
Since the external and internal chord diagrams together determine the 
number of closed loops between them, this is equivalent to comparing 
the partition functions of the two lattices, which are identical due to 
the $Z$-invariance guaranteed by the integrability conditions already 
proven. Note that even though not all of the configurations in the 
definition of the 
observable in the dilute loop model is allowed in the partition 
function, the internal configurations are all allowed since the open 
path ends on its boundary.

We thus have the invariance of the observable on the boundary under
Reidemeister moves of the rapidity lines inside, or equivalently, 
local reshuffling of rhombuses in the embedding. This provides another
characterization of the observables besides discrete holomorphicity.
It is interesting to note that discrete holomorphicity then provides
linear relations among partition functions with different chord 
diagrams on the same lattice, whereas the integrability conditions 
show the equality of partition functions with the same chord diagram
on different lattices. Of course, for just one rhombus, the partition
functions with different chord diagrams are just the Boltzmann weights
themselves.

\section{Conclusion}
In summary, discrete holomorphicity of the observables, on 
simply-connected domains tiled by convex rhombuses, generates linear 
equations between the partition functions with different internal 
connectivities of the
boundary points. For one rhombus, these are linear equations in the 
weights, which are solved by finding the conformal spin by setting
the determinant of the system of equations to zero. 
For two rhombuses, these equations are 
quadradic in the weights, and imply the inversion relation. For three
rhombuses, they imply the Yang-Baxter relation. In general, for a 
domain with $k$ rhombuses, the linear equation is among degree $k$
homogeneous polynomials in the weights. With the integrability 
conditions satisfied, the $Z$-invariance of the models defines the 
observable on the boundary of the domain for the equivalence class
of Baxter lattices.

We note, however, that unlike the case of Cauchy's integral formula, 
\[
\psi(z_0) = 
\frac{1}{2\pi\ii}\oint_{\partial\Omega} \frac{\psi(z)}{z-z_0}\;dz
\]
the values of the observable on the boundary do not uniquely specify
the values in the interior. As was explained in \cite{C-DHCP}, this is
because the number of variables, that is the number of midpoints of
the sides of the rhombuses, far exceeds the number of equations, that
is, one equation of holomophicity per rhombus. The observable defined
on the Ising model \cite{SS-CIRCM}, of course, is the glaring 
exception.

\ack We thank John Cardy for his invaluable guidance and kind
hospitality in Oxford. We also thank Vladimir Bazhanov
for his inspirations and insights, and Yvan Saint-Aubin,
Jan de Gier, Alexander Lee and Rodney Baxter for helpful
discussions. We thank the Rudolf Peierls Centre for Theoretical Physics 
for their hospitality during our visits to Oxford. MTB particularly 
thanks All Souls College for support through a Visiting Fellowship. 
He also acknowledges support from the 1000 Talents Program of China and 
from Chongqing University. This work has also been supported by the 
Australian Research Council through grant DP130102839.

\appendix
\setcounter{section}{1}
\section*{Appendix}
In this section, we show that the vanishing differences of the 
discrete contour sums around the hexagon with the two arrangements, 
the star and the triangle, for the 21 different connectivities, imply
the Yang-Baxter equations for the dilute loop model discussed in
\sref{sec:DiluteYangBaxter}. Of the equations, the following subset
suffices,
\begin{eqnarray}
\fl\phi(\alpha) \wrap{\mathrm{YB}_4(\alpha,\gamma,\beta)
-n \mathrm{YB}_4(\beta,\gamma,\alpha)}
+n \mathrm{YB}_2(\alpha,\beta,\gamma)
-\phi(\pi)\mathrm{YB}_1(\alpha,\gamma,\beta)\nonumber\\
+\;\phi(\alpha+\pi) \mathrm{YB}_3(\gamma,\alpha,\beta)
-\phi(\alpha-3 \pi) \mathrm{YB}_5(\alpha,\gamma,\beta) = 0
\label{eq:eq3}\\
\fl\phi(\alpha -\pi) \left(\mathrm{YB}_4(\beta,\gamma,\alpha)
-n \mathrm{YB}_4(\alpha,\gamma,\beta)\right)
+\phi (\alpha) \left(\mathrm{YB}_5(\alpha,\gamma,\beta)
-n \mathrm{YB}_3(\gamma,\alpha,\beta)\right)\nonumber\\
+\;n \mathrm{YB}_1(\alpha,\gamma,\beta )
+\phi(-\pi) \mathrm{YB}_2(\gamma,\beta,\alpha) = 0
\label{eq:eq7}\\
\fl\phi(-\beta) \mathrm{YB}_1(\alpha,\gamma,\beta)
-\phi(\alpha) \mathrm{YB}_1(\beta,\gamma,\alpha)
+\phi(\pi -\beta) \mathrm{YB}_2(\alpha,\beta,\gamma)\nonumber\\
+
\;\phi(\alpha -\pi) \mathrm{YB}_2(\gamma,\alpha,\beta ) = 0
\label{eq:eq1}\\
\fl\phi(\pi -\beta) \left(n \mathrm{YB}_4(\beta,\gamma,\alpha )
-\mathrm{YB}_4(\alpha,\gamma,\beta)\right)
+\phi(-\beta) \left(n \mathrm{YB}_3(\gamma,\alpha,\beta)
-\mathrm{YB}_5(\alpha,\gamma,\beta )\right)\nonumber\\
-\;n \mathrm{YB}_1(\beta,\gamma,\alpha )
+\phi(\pi) \mathrm{YB}_2(\beta,\alpha,\gamma ) = 0
\label{eq:eq4}\\
\fl\phi (-\beta) \left(n \mathrm{YB}_4(\alpha,\gamma,\beta)
-\mathrm{YB}_4(\beta,\gamma,\alpha)\right)
+n \mathrm{YB}_2(\gamma,\alpha,\beta )
+\phi(-\pi) \mathrm{YB}_1(\beta,\gamma,\alpha)\nonumber\\
-\;\phi (-\beta -\pi) \mathrm{YB}_3(\gamma,\alpha,\beta)
+\phi (3 \pi -\beta) \mathrm{YB}_5(\alpha,\gamma,\beta ) = 0
 \label{eq:eq12}\\
\fl
\phi(\alpha) \left(1-n^2 \right)
\mathrm{YB}_6(\alpha,\beta,\gamma)
+n^2 \mathrm{YB}_3(\alpha,\beta,\gamma)
-n \mathrm{YB}_5(\beta,\alpha,\gamma)\nonumber\\
+\;\phi(2 \pi -\beta)\left(n\mathrm{YB}_5(\alpha,\gamma,\beta)
-\mathrm{YB}_3(\gamma,\alpha,\beta)\right)\nonumber\\
+\;\phi (-\beta -\pi) \left(n \mathrm{YB}_4(\alpha,\gamma,\beta)
-\mathrm{YB}_4(\beta,\gamma,\alpha)\right)\nonumber\\
+\;\phi (-\pi) \left(n \mathrm{YB}_4(\beta,\alpha,\gamma)
-\mathrm{YB}_4(\gamma,\alpha,\beta)\right) = 0
\label{eq:eq8}
\end{eqnarray}

In addition, to utilize the symmetries of the $\mathrm{YB}_i$ functions,
we make the following replacements whenever the left hand sides occur
in the calculations,
\begin{eqnarray*}
\mathrm{YB}_2(\gamma,\beta,\alpha) \to
-\mathrm{YB}_2(\alpha,\beta,\gamma) \quad\quad& 
\mathrm{YB}_2(\gamma,\alpha,\beta)\to 
-\mathrm{YB}_2(\beta,\alpha,\gamma )\\
\mathrm{YB}_5(\beta,\gamma,\alpha)\to 
\mathrm{YB}_5(\alpha,\gamma,\beta) \quad\quad&
\mathrm{YB}_3(\gamma,\beta,\alpha)\to 
\mathrm{YB}_3(\gamma,\alpha,\beta)
\end{eqnarray*}

We ignore isolated points for the values of $n$ to keep the proof 
general, although the special cases can be handled similarly, or
using the continuity of the $\mathrm{YB}$ functions and the 
weights. We start by multiplying equation \eref{eq:eq3} by $\phi(-\pi)$ 
to express $\mathrm{YB}_1$ in terms of the others, enabling us to 
eliminate its subsequent occurences,
\begin{eqnarray*}
\fl\mathrm{YB}_1(\alpha,\gamma,\beta) = 
n \phi(-\pi) \mathrm{YB}_2(\alpha,\beta,\gamma)
-n \phi (\alpha -\pi) \mathrm{YB}_4(\beta,\gamma,\alpha)\\
+\;\phi(\alpha) \mathrm{YB}_3(\gamma,\alpha,\beta )
+\phi(\alpha -\pi) \mathrm{YB}_4(\alpha,\gamma,\beta)
-\phi(\alpha - 4 \pi) \mathrm{YB}_5(\alpha,\gamma,\beta)
\end{eqnarray*}
so that equation \eref{eq:eq7} gives, after multiplication by 
${\phi (\pi -\alpha )}/\wrap{n^2-1}$,
an expression for $\mathrm{YB}_4$,
\begin{eqnarray*}
\fl\mathrm{YB}_4(\beta,\gamma,\alpha) = \frac{1}{n^2 - 1}
\Big(n^2 \phi(-\alpha) \mathrm{YB}_2(\alpha,\beta,\gamma)
+ \phi(-\alpha ) \mathrm{YB}_2(\gamma,\beta,\alpha)\\
-\;(n \phi(-3 \pi)-\phi(\pi)) \mathrm{YB}_5(\alpha,\gamma,\beta)
   \Big)
\end{eqnarray*}

We now multiply by $(n + 1)$ and substitute $\mathrm{YB}_1$ and 
$\mathrm{YB}_4$ into equations \eref{eq:eq1}, \eref{eq:eq4}, and
\eref{eq:eq12} successively, to get
\begin{eqnarray}
\label{eq:simple1}
\fl 
(-\phi(\alpha - \beta)-\phi(\alpha -\beta -4 \pi)
+ \phi(\alpha + \beta) + \phi (\alpha +\beta -4 \pi )) 
\mathrm{YB}_5(\alpha,\gamma,\beta)
\nonumber\\
-\;(n+1) \wrap{\phi(\alpha -\pi)-\phi(\alpha - 2 \beta - \pi)} 
\mathrm{YB}_2(\beta,\alpha,\gamma)\nonumber\\
+\; (n+1) \wrap{\phi(\pi -\beta)-\phi(\beta-\pi)} 
\mathrm{YB}_2(\alpha,\beta,\gamma) \nonumber\\
+ \;(n+1) (\phi (\alpha -\beta)-\phi (\alpha +\beta)) 
\mathrm{YB}_3(\gamma,\alpha,\beta) = 0
\end{eqnarray}
\begin{eqnarray}
\label{eq:simple4}
\fl(\phi (2 \pi -\beta)-\phi(-\beta)-n \phi (-\beta -2 \pi)
-n \phi(-\beta) + n \phi(\beta) + n \phi (\beta - 4 \pi )) 
\mathrm{YB}_5(\alpha,\gamma,\beta)\nonumber\\
+\;(n+1) (\phi (\pi)-\phi (\pi -2 \beta)) 
\mathrm{YB}_2(\beta,\alpha,\gamma)\nonumber\\
+\;n (n+1) (\phi (-\beta)-\phi (\beta)) 
\mathrm{YB}_3(\gamma,\alpha,\beta)\nonumber\\
+\; n (n+1) (\phi(-\alpha -\beta +\pi )-\phi(-\alpha +\beta -\pi)) 
\mathrm{YB}_2(\alpha,\beta,\gamma) = 0
\end{eqnarray}
\begin{eqnarray}
\label{eq:simple12}
\fl
\wrap{\phi(-\beta)\wrap{\phi (\pi)+\phi(3 \pi)
-n \phi (-3 \pi)+n \phi (3 \pi)}-\phi(\beta)\wrap{\phi(-5 \pi)
+\phi(-\pi)}} 
\mathrm{YB}_5(\alpha,\gamma,\beta) \nonumber\\
+\;n (n+1) (\phi (-2 \beta)-1) \mathrm{YB}_2(\beta,\alpha,\gamma)
\nonumber\\
-\;(n+1) (\phi (-\beta -\pi)-\phi (\beta -\pi)) 
\mathrm{YB}_3(\gamma,\alpha,\beta )\nonumber\\
-(n+1) (\phi (-\alpha -\beta )-\phi (-\alpha +\beta -2 \pi)) 
\mathrm{YB}_2(\alpha,\beta,\gamma) = 0
\end{eqnarray}

Then we eliminate $\mathrm{YB}_3$ from (\ref{eq:simple1}) 
and (\ref{eq:simple4}) to obtain
\begin{eqnarray}
\label{eq:comb1and4}
\fl(n+1) (\phi(\pi) + n \phi(-\pi))(\phi(-\beta) - \phi(\beta))
\mathrm{YB}_2(\beta ,\alpha ,\gamma )\nonumber\\
+\;(-n \phi (-4 \pi)+n \phi (-2 \pi)-\phi(2 \pi)+1) 
\mathrm{YB}_5(\alpha,\gamma,\beta) = 0
\end{eqnarray}
whereas eliminating $\mathrm{YB}_3$ from (\ref{eq:simple4}) 
and (\ref{eq:simple12}) gives
\begin{eqnarray}
\label{eq:comb4and12}
\fl(n-1)(n+1) (\phi (-\beta)-\phi (\beta)) 
\mathrm{YB}_2(\beta,\alpha,\gamma)\nonumber\\
+\;(-n \phi (-3 \pi)+n \phi (3 \pi)-\phi (-\pi)+\phi (\pi))
\mathrm{YB}_5(\alpha,\gamma,\beta) = 0
\end{eqnarray}

From (\ref{eq:comb1and4}) and (\ref{eq:comb4and12}) we can eliminate
$\mathrm{YB}_2$,
\begin{equation}
\fl (n \phi (-2 \pi)-n \phi (2 \pi)+\phi (-4 \pi)+\phi (-2 \pi)
-\phi (2 \pi)-\phi (4 \pi)) \mathrm{YB}_5(\alpha,\gamma,\beta) = 0
\end{equation}
Since the prefactor is not identically zero, 
\begin{equation*}
\mathrm{YB}_5(\alpha,\beta,\gamma) = 0
\end{equation*}

Back-substituting, in succession, into  (\ref{eq:comb4and12}), 
(\ref{eq:simple4}), the expressions for $\mathrm{YB}_4$ and 
$\mathrm{YB}_1$, and \eref{eq:eq8}, we successively obtain
\begin{eqnarray*}
\mathrm{YB}_2(\alpha,\beta,\gamma) = 0\\
\mathrm{YB}_3(\alpha,\beta,\gamma) = 0\\
\mathrm{YB}_4(\alpha,\beta,\gamma) = 0\\
\mathrm{YB}_1(\alpha,\beta,\gamma) = 0\\
\mathrm{YB}_6(\alpha,\beta,\gamma) = 0
\end{eqnarray*}
that is, the Yang-Baxter equations are satisfied.

\end{document}